\documentstyle[epsf,a4]{article}  
\begin{document}                
\newcommand{\manual}{rm}        
\newcommand\bs{\char '134 }     
\newcommand{\Het}{$^3{\mathrm{He}}$}
\newcommand{\Hef}{$^4{\mathrm{He}}$}
\newcommand{\A}{{\mathrm{A}}}
\newcommand{\D}{{\mathrm{D}}}
\newcommand{\simlt}{\stackrel{<}{{}_\sim}}
\newcommand{\simgt}{\stackrel{>}{{}_\sim}}
\newcommand{\MeV}{\;\mathrm{MeV}}
\newcommand{\TeV}{\;\mathrm{TeV}}
\newcommand{\GeV}{\;\mathrm{GeV}}
\newcommand{\eV}{\;\mathrm{eV}}
\newcommand{\cm}{\;\mathrm{cm}}
\newcommand{\s}{\;\mathrm{s}}
\newcommand{\sr}{\;\mathrm{sr}}
\newcommand{\lab}{\mathrm{lab}}
\newcommand{\ts}{\textstyle}
\renewcommand{\floatpagefraction}{1.}
\renewcommand{\topfraction}{1.}
\renewcommand{\bottomfraction}{1.}
\renewcommand{\textfraction}{0.}               
\renewcommand{\thefootnote}{F\arabic{footnote}}
\title{Are neutral Goldstone bosons initiating very energetic air showers
and anomalous multiple-core structure as a component of cosmic rays?}
\author{Saul Barshay and Georg Kreyerhoff\\
III. Physikalisches Institut\\
RWTH Aachen\\D-52056 Aachen\\Germany}
%
\maketitle
\begin{abstract}                
We consider two recently accentuated, unusual empirical results concerning 
cosmic-ray events at high energies. We show that the possibility for
a correlated explanation is provided by new dynamics which arises from
collisions of a neutral Goldstone boson as a component of the highest-energy
cosmic rays.
\end{abstract}
In this paper, we consider the hypothesis that there is a nearly-massless
neutral Goldstone boson, $b^0$, which is a component of high-energy
cosmic rays and which is initiating air showers \cite{ref1,ref2,ref3} at
the highest energies and possibly also anomalous multiple-core structure
seen in photo-emulsion chamber experiments \cite{ref4,ref5,ref6,ref7,ref8}.
We point out and examine the consequences of new dynamics which arises
from the assumption that there is an effective interaction originating
in a neutral triangle anomaly \cite{ref9,ref10} involving $b^0$ and a
neutral vector boson $Z'$ (and/or a chirally-related neutral axial vector 
boson). This possibility may be related to the possibility that neutrino
mass originates as a consequence of a spontaneously broken chiral symmetry
at a low energy scale, $F < 0.4\MeV$. The strength of the effective
interaction is $(\frac{g^2}{2\pi^2F})$ with $g^2$ assumed, for numerical
illustration of the idea, to be of the order of 0.1 (i.~e.~like a
strength for $Z$).\footnote{The vertex arising from the triangle anomaly leads to
higher-order nonrenormalizable divergences, so the model must involve
effective interactions that are restrained at some high energy scale.
Summing (strong) ``bubble'' diagrams might restrain contributions
to $m_b^2$ i.~e.~$\delta m_b^2(p^2)\sim (m_b^2-p^2)$ at squared four-momentum
$p^2$.} The communication of $b^0$ with
quarks is only via $Z'$ exchange; among leptons, $Z'$ likely couples only to 
neutrinos (like $b^0$).Then the mass of $Z'$ may not be very much greater
than that of $Z$. An actual value up to some hundreds of GeV does not
affect our display of interesting dynamics, because we are considering
cosmic-ray collisions at (c.~m.) energies well above $m_{Z'}$, 
i.~e.~$>10\TeV$ for $b^0$-nucleon collisions ($>1\TeV$ for
an idealized $b^0$-quark collision).\par
Our main purpose is to clearly exhibit the relevance of the hypothesis
to providing new dynamics bearing upon two long-standing puzzling aspects of
cosmic-ray interactions at the highest energies.
1. The presence of a few ``hadron-like'' air shower events at primary
energies estimated to be about $10^{20}\eV$. \cite{ref1,ref2} The data
still is limited, but this is an energy above the Greisen-Zatsepin-Kuzmin
cut-off \cite{ref12,ref13} for arrival of protons from sources at very
great (cosmological) distances, because of their interaction with the
cosmic microwave background photons. These events could be related
to neutral $b^0$ arriving from even the largest distances and interacting
rather strongly in the atmosphere, if a sufficient flux of $b^0$ at $\sim
10^{20}\eV$ is attainable. We estimate below that a flux of $\sim
10^{-20}(\cm^2-\s-\sr)^{-1}$ is attainable for $b^0$ originating in the decay
of massive inflatons which we have shown \cite{ref14} can constitute much
of dark-matter, and for which we have calculated \cite{ref14} a lifetime
orders of magnitude greater than the present age of the universe.
2. Over many years, a number of unusual events of a particular kind 
\cite{ref6,ref7} at high energies have been observed in several photo-emulsion
chamber experiments at different high-altitude locations \cite{ref7}.
These events at $E_{\lab}\sim 10^{16}-10^{17}\eV$ involve double-core
$\gamma$-families, where a $\gamma$-family consists of a bundle of high-energy
particles incident on the chamber in almost the same direction. In the 
laboratory, the direction is in the very forward region for the collisions.
Recently \cite{ref6,ref7,ref8}, these events have been analyzed in terms
of high transverse momentum jet production in QCD. The result of this
analysis \cite{ref6,ref7,ref8}, which must be viewed in the context
of limited statistics, is stated to be an order of magnitude excess of
events at the largest inferred \cite{ref8} transverse momenta.  It is said
to be suggestive of new physics above $\sim 10^{16}\eV$ characterized
by particle production in the forward direction with unusually large
transverse momenta. \cite{ref8} \footnote{When discussing unusual behavior in the forward direction,
it is important to recognize that the cross section for single diffractive
dissociation must tend to zero at extremely high energies, as the interaction
of hadrons approaches that of black disks. See \cite{ref15}} 
In fact, an early summary of cosmic 
rays \cite{ref4} already presented a number of events involving 
``binocular'' families, analysis of which indicated an unusual general
characteristic of forward production at transverse momenta of some
tens of GeV of a decaying entity with an effective mass of tens of GeV.
Subsequent data \cite{ref5} indicated family events at energies above 
$10^{16}\eV$ characterized by large production transverse momenta,
and a rich hadron content. We show below that the cross section for
production of $Z'$ by $b^0$ must be dynamically enhanced at the largest
values of the squared four-momentum transfer to a nuclear target. There
is  a large deep-inelastic cross section. Decay of $Z'$ into quark
and antiquark and quark ejection from the target (followed by fragmentation
and secondary hadronic interactions) naturally gives rise to events 
having two cores with large transverse momenta in forward directions
(and also possibly to three such cores \cite{ref4}).\par
To clearly see the dynamical features, consider production of $Z'$ 
in a $b^0$-quark collision via exchange of $Z'$, as in the diagram
in Fig.~1a. The interaction at the upper vertex is $(\frac{g^2}{2\pi^2F})
(\frac{1}{4}\epsilon_{\mu\nu\sigma\rho}F_{Z'}^{\mu\nu}F_{Z'}^{\sigma\rho})$;
at the lower vertex $g(\overline{\psi}_q \gamma^\mu\psi_qZ'_\mu)$. 
The differential cross section is then
\begin{equation}
\frac{d\sigma}{d(\cos\theta)}\cong \left(\frac{g^2}{2\pi}\right)
\left(\frac{g^2}{2\pi^2F}\right)^2\frac{ \frac{\ts s^2}{\ts 4} \left(
\sin^2\theta + \frac{1}{4}(1-\cos\theta)^3\right)}
{\left(\frac{\ts s}{2}(1-\cos\theta)+m_{Z'}^2\right)^2}
\end{equation}
The mass $m_{Z'}$ (taken as $\sim m_Z$ simply for numerical illustration) is
retained only in the propagator (to avoid singular behavior in the squared
four-momentum transfer $-t =\frac{\ts s}{\ts 2}(1-\cos\theta)$). For c.~m.~energy
$\sqrt{s}\sim 1\TeV$ ($E_{\lab}^b\sim 10^{17}\eV$ on $m_q\sim 5\MeV$ ), the total
cross section is large, $> 10\;{\mathrm{mb}}$. This is the first feature, which is
due to the low scale $F$. The second feature is the marked enhancement
at the largest $|t|$, which is caused by the numerator function in Eq.~(1).
\footnote{ This behavior is also present in the cross section for the inverse
Primakoff effect, which involves production of a photon by an incident pseudoscalar
boson in a nuclear Coloumb field. See \cite{ref16}}  
Most of $\sigma$ arises from $|t|\ge m_{Z'}^2$. These two
features carry over to the realistic situation involving deep-inelastic
interactions of $b^0$ impinging upon nucleons in the atmosphere, as in the
diagram in Fig.~1b. In terms of the usual parton model variables 
$x=\frac{|t|}{2m_N(E_b-E_{Z'})}$, $y=\frac{(E_b-E_{Z'})}{E_b}$, and
the nucleon structure function $F_2(x,|t|)$, the laboratory differential
cross section is
\begin{equation}
\frac{d^2\sigma_b}{dxdy}=\left(\frac{g^2}{4\pi}\right)\left(\frac{g^2}
{2\pi^2F}\right)^2\frac{\frac{\ts s}{\ts 8}|t|}{(|t|+m_{Z'}^2)^2}\frac{F_2}{2  }
\left(1+(1-y)^2+\frac{y^2}{2}\right)
\end{equation}
with $s\cong2m_N p_b$, $|t|\cong 4p_bp_{Z'}\sin^2\frac{\theta}{2}$.
We have retained $m_{Z'}$ only in the propagator, $m_N$ is the nucleon mass,
$p_b$, $p_{Z'}$ are momenta, and $\theta$ is the angle of $Z'$. The
total cross section is greater than $10^{-2}$ mb. This is a large
deep-inelastic cross section, because it is controlled by $\left(
\frac{1}{2\pi^2F}\right)^2$. Compare Eq.~(2) to the usual cross section
for deep-inelastic scattering of a charged lepton via photon exchange,
whose size is controlled by $\sim\frac{1}{|t|}$
\begin{equation}
\frac{d^2\sigma_\ell}{dxdy}=\frac{(e^2)^2}{4\pi}\frac{s}{(|t|^2)^2}
\frac{F_2}{2}\left(1+(1-y)^2\right)
\end{equation}
In this comparison, one sees again the enhancement at the largest
$|t|$ due to the numerator in Eq.~(2). The double-core events do
indicate a fall off consistent with $(\frac{1}{p_T^2})$, not 
$(\frac{1}{p_T^4})$, at the largest inferred \cite{ref8} transverse
momenta $p_T$. The two features are just what is required to deal
with the two unusual empirical aspects of the high-energy cosmic-ray
interactions, if sufficient flux of $b^0$ here is achievable.\par
Consider first $b^0$ near to the maximum energy, which we consider
to be $\sim 10^{20}\eV$. The maximum energy of $b^0$ arises from their 
possible origin in the decay of inflatons $\phi$, $\phi\to bb$. We
have calculated the $\phi$ mass in detail \cite{ref14} in
a specific phenomenological cosmological model. It is in a limited range
$\sim 10^{19} - \sim 10^{20}\eV$. The mass range is based upon explicit
calculation of inflaton potentials which exhibit \underline{both}
a minimum at an energy scale $\phi_c\cong 10^{16}-10^{17}\GeV$ and
a maximum at the Planck scale. \cite{ref14,ref17} We have shown
that the massive inflaton quanta produced near the end of inflation
can constitute much of dark matter today, with an energy density
estimated to be in the range $0.1-0.5$ of critical. The inflaton may
not be completely stable. In the model \cite{ref14}, it has no coupling
to ordinary matter except for a possible very weak coupling to neutrinos
proportional to neutrino mass; this leads to the decay $\phi\to
\nu_\tau\overline{\nu}_\tau$ with a lifetime given by\footnote{
The factor $\sqrt{10^{-5}}\sim \left(\frac{1}{30\pi^2}\right)$ arises
from a (conservative) estimate of the contribution to the model matrix element from
a neutral, massive intermediate state which ``mixes'' into $\nu_\tau\overline{\nu}_\tau$
with an effective coupling proportional to $(\sqrt{m_{\nu_\tau}/\phi_c})^2$.}
\begin{equation}
\tau_\phi=(\Gamma_\phi)^{-1}\sim\left(\left(10^{-5}\right)
\left(\frac{m_{\nu_\tau}}{\phi_c}\right)^2\frac{m_\phi}{8\pi}\right)^{-1}
\cong 10^{26}\s
\end{equation}
where we use here $m_{\nu_\tau}\cong 0.06\eV$ as a hypothetical 
heaviest-neutrino mass, and our calculated \cite{ref14} values 
$\phi_c\cong 10^{17}\GeV$, $m_\phi\cong5\times 10^{10}\GeV$.
A possible ``mixing'' between the chiral-like dynamics\footnote
{ The chiral dynamics suggests a reason for the necessary smallness
of the inflaton quartic self-coupling, $\lambda\sim 10^{-13}$ which leads to $m_\phi\cong             \sqrt{\lambda}\phi_c\sim 5\times 10^{10}\GeV$.}
which we examined in the cosmological model \cite{ref14},
and a possible low-scale chiral dynamics involved in neutrino
mass suggests an effective interaction $f(\phi b b)$, with the order
of magnitude of f given by\footnote{In a chiral $\sigma$-model, $b^0$ couples to 
neutrinos with strength $(m_\nu/F)$, i.~e.~note \cite{ref20}}
\begin{equation}
f\sim \left(\sqrt{10^{-5}}m_{\nu_\tau}\right)\left(\frac{m_{\nu_\tau}}
{4\pi^2F}\right)
\end{equation}
This results in a definite branching ratio, $(\Gamma(\phi\to bb)/
\Gamma(\phi\to\nu_\tau\overline{\nu}_\tau))=r\sim 0.5\times 10^{-4}$.
We have estimated \cite{ref21,ref22,ref14} that the decay of dark matter
can give rise to significant flux of the maximum-energy neutrinos
here, about $2\times 10^{-16}(\cm^2-\s-\sr)^{-1}$ from within
our galaxy alone. The estimate for $r$ thus implies a flux of $b^0$ of
$\sim 10^{-20}(\cm^2-\s-\sr)^{-1}$. In fact, the recent AGASA data \cite{ref1}
consists of a handful of ``hadron-like'' events, all close to
$10^{20}\eV$, in an exposure of about $2.6\times 10^{20}(\cm^2-\s-\sr)$.
We have given numerical examples of a possible ``bump'' structure \cite{ref22}
in events near to the maximum energy, that is a structure above the GZK
cut-off and near to $10^{20}\eV$.\par
The anomalous double-core events \cite{ref8} imply a flux of $b^0$ of the
order of $10^{-16}(\cm^2-\s-\sr)^{-1}$ at say $\sim 10^{16.5}\eV$. These
could be produced from the inverse of the process shown in Fig.~1a.
The $Z'$ would have to be produced in exceedingly dense matter in order
to interact before decay. The unique possibility would seem to be the 
densities reached by neutron-star matter. This suggests neutron stars as
discrete sources for at least some of the unusual high-energy cosmic
rays. In this respect, it is worth noting that there is recent 
evidence \cite{ref3} for anisotropy in the arrival directions for cosmic rays 
around $10^{18}\eV$, with a significant excess near the directions
of the galactic center and the Cygnus region \cite{ref3}. The excess
flux is at the level of $\sim 10^{-18}(\cm^2-\s-\sr)^{-1}$. Among
the events above $10^{19}\eV$, there is a curious tendency to
clustering of directions \cite{ref2}. One significant cluster is 
said \cite{ref2} to encompass the direction of a known pulsar, as well as
that of the Cygnus ``loop''. Of course, protons may be accelerated
to high energies near magnetized neutron stars.\footnote{
If produced from quarks at accelerator energies, $Z'$ would give rise
to some ``extra'' hadronic jets, and possibly to some missing energy in neutrinos.}
\footnote{The existence of anisotropy restricted to a narrow range of energies \cite{ref3}
is suggestive of production of neutral particles by higher-energy particles at discrete
sources. If these particles are concentrated at a definite energy (as in two-body
decay products of a massive particle), then the particular produced particle will
have its ``typical'' energies cut-off both at high values {\underline {and}} at
low values by the low and high-multiplicity tails of multiplicity distributions,
respectively.}
\par
The ideas in this paper illustrate the fact that the absence of events
characterized by new dynamics at the LHC, for $pp$ collisions at c.~m.~energies
$\sim 10\TeV$, would not refute the presence of new phenomena in cosmic-ray
events at $E_{\lab}>10^{16}\eV$. Evidently, the particular nature of an
incident particle can be crucial, as well as the energy which it
brings into the collision process. The corollary is that the cosmic-ray
experiments and the coming experiments sensitive to fluxes of very high 
energy neutrinos (above $10^{19}\eV$) are essential \cite{ref24}.
The comment in this last paragraph could apply also to cosmic-ray
Centauro (and anti-)events, not detected at present accelerators.
Coalescence of the final-state $Z'$ and an atmospheric target excited to mass
of $\sim m_{Z'}$ might be a uniquely favorable dynamical configuration for the
transient formation of a coherent state of pions \cite{ref26}. This state
could result in Centauro-type characteristics with $\langle n_\pi\rangle
\sim 2\langle n_\pi\rangle \simgt 2\langle n_\pi\rangle_Z \simgt 56$, and
$\langle p_T^2 \rangle ^{1/2} \sim \sqrt{\frac{2}{3}}
\left(\frac{m_{Z'}}{\langle n_\pi\rangle_{Z'}}\right)\sim \sqrt{\frac{2}{3}} 
\left(\frac{m_Z}{\langle n_\pi\rangle_{Z}}\right)\cong 2.6 (\mathrm{GeV}/c)$.
The empirical numbers for Centauro I at primary energy of $\sim 1650$ TeV, are
$\sim 74$ and $ \sim 1.8 (\mathrm{GeV}/c)$), respectivly \cite{ref27}.

S.~B.~is grateful to Dr.~Patrick Heiliger for much help.

\newpage
\section*{Figures}
\renewcommand{\thefigure}{1\alph{figure}}
\begin{figure}[h]
\begin{center}
\mbox{\epsfysize 8cm\epsffile{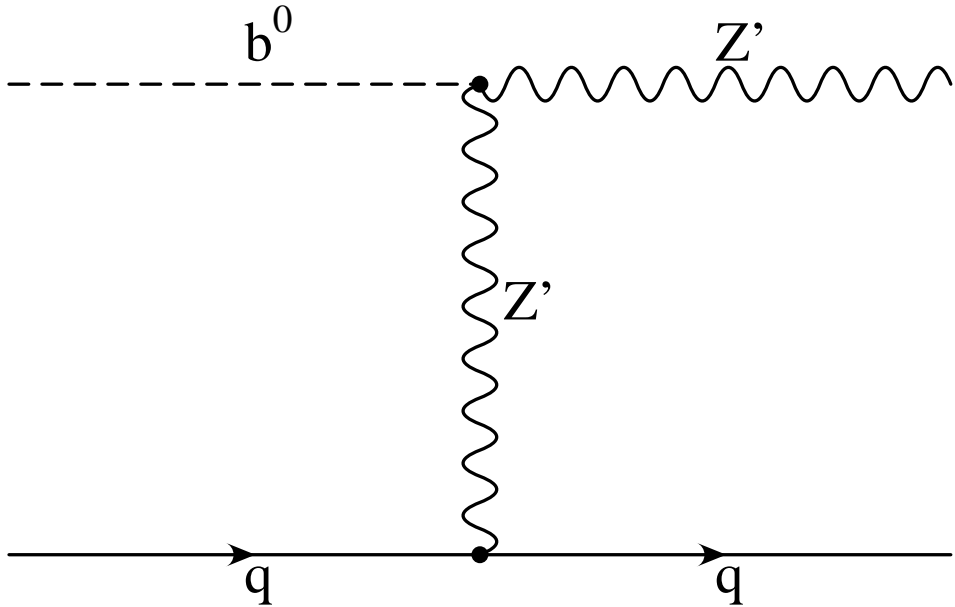}}
\caption{Production of a neutral massive vector boson $Z'$ by a pseudoscalar
Goldstone boson $b^0$ in collision with a quark $q$.}
\end{center}
\end{figure}
\begin{figure}[b]
\begin{center}
\mbox{\epsfysize 8cm\epsffile{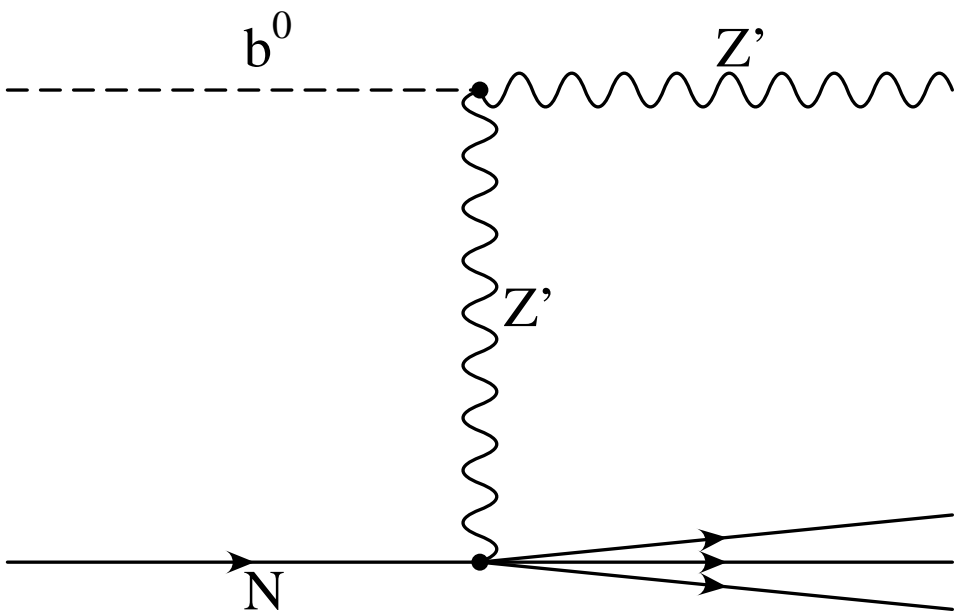}}
\caption{Deep-inelastic scattering initiated by $b^0$ in collision with a nucleon $N$.}
\end{center}
\end{figure}
\end{document}